\documentclass[preprint, amsmath]{revtex4}
\usepackage{amssymb}
\usepackage{indentfirst}
\usepackage[mathscr]{eucal}
\usepackage{graphicx}

\begin{document}

\title{
        Polarons in the harmonic lattice. \\
        I. Standing polaron.
       }

\author{
         V.N. Likhachev, T.Yu. Astakhova, G.A. Vinoghradov
        }

\affiliation{
 Emanuel Institute of Biochemical Physics, Russian Academy of
 Sciences, \\
 ul. Kosygina 4, Moscow 119991, Russia
             }

\email{gvin@deom.chph.ras.ru}

\begin{abstract}

We obtain analytical expressions for the large- and small-radius
polarons on the one-dimensional lattice in the TBA
approximation. The equations of motion for this model are
treated classically for the oscillator subsystem, while a
quantum description is used for the electron. The
electron-phonon interaction is considered in the linear
Su--Schrieffer--Heeger approximation. Good agreement between
analytical formulae and accurate numerical simulation is
obtained. The dynamics of polaron formation from different
initial conditions is considered. Some features of the wave
function evolution, governed by the finite lattice length, are
elucidated.

\vspace{0.5 cm}

\noindent{\it Keywords }: polaron, DNA, charge transfer

\noindent PACS numbers: 71.38.-k, 87.14.Gg, 87.15.-v

\end{abstract}

\maketitle



\section{Introduction}%
  \label{sec:Intro}

Charge and energy transfer is of utmost importance in both
organic and inorganic nature. Studies of charge transport (CT)
in organic systems were restricted till recently by research of
amorphous state. CT in individual quasi-one-dimensional systems
has attracted much attention. These oligo-- and macromolecules
have some advantages compared to traditional bulk systems
because of less size, flexibility and power requirements.
Biopolymers like natural and synthetic DNA, and proteins
\cite{Del03, Gen10a, Mal10, Shi10} are of particular interest.
These systems can find an applications in solar cells, flexible
TV displays, logical circuits etc. \cite{Wag05, Mal05, Rob03,
Che09, Son09}. Molecular electronics can also be used in the
nanotechnology   \cite{Off09, Lin07, Shi08, Pau08, Hat08, Pau09,
Fuj04}. Photosynthesis, ATP hydrolysis, metabolism are examples
where CT is vitally important \cite{Coo03, Loo10, Dav07}.

Polarons are ubiquitous in materials where the electron-phonon
coupling cannot be ignored. Polarons are thought to be
responsible for the CT in biological systems \cite{Con00, Con01,
Con05} including DNA. Different DNA models were considered to
describe the polaron properties \cite{Kal98, Kuc10, Sin10,
Dev09, Su_79}. Dozens of papers present results on the effects
of water and counterions surrounding DNA, on polaron properties
in the presence of disorder, polaron hopping and drift in an
applied electric field \cite{Kuc10, Zhe06, Sch08, Ale03, Con04,
Zha04, Con07, Wei08, Mah10, Kal04, Hen99, Tri06, Li_09}.

Polarons in the harmonic lattice are thoroughly investigated in
the resent paper. The electron-phonon interaction is described
in terms of the linear Su--Schrieffer--Heeger approximation in
the tight-binding modelling. Analytical expressions for small
and large polarons are derived. Results coincide with numerical
simulation with high accuracy. The dynamics of polaron formation
from different initial conditions is considered. An influence of
the lattice finite size on the wave function evolution is also
analyzed.


\section{The small-radius polaron in the harmonic lattice}
  \label{sec:small_polaron}

\subsection{Numerical simulation}

The one-dimensional harmonic lattice with one extra charge
carrier is investigated. We consider an electron for the
definiteness though the arguing is valid for the holes. The
lattice consists of $N$ particles. The "particle" represents one
base of DNA in the coarse-grained model of DNA. The TBA
approximation is used to describe the electron evolution on the
lattice.

The hamiltonian of this system is the sum of the classical
hamiltonian and the energy of electron-phonon interaction:
\begin{equation}
  \label{a}
 H = \dfrac{m}{2}\sum\limits_{i=1}^N \dot x_i^2 +
     \dfrac{k}{2}\sum\limits_{i=1}^{N-1} (x_{i+1}-x_i)^2 +
     \left< \vec\Psi | \widehat H^{\rm e} | \vec\Psi \right> \,,
\end{equation}
where $x_i$ is the deviation of $i$-th particle from the
equilibrium, $\vec\Psi = \psi_1, \psi_2, \ldots$ is the electron
wave function. $m$ and $k$ are the particle mass and the lattice
rigidity, respectively. In the TBA approximation the electron
hamiltonian is the tridiagonal matrix:
\begin{equation}
 \widehat H^{\rm e} =
 \begin{pmatrix}
  \label{b}
 e_1    & t_1    & 0      & 0      & \ldots & 0       & 0      \\
 t_1    & e_2    & t_2    & 0      & \ldots & 0       & 0      \\
 0      & t_2    & e_3    & t_3    & \ldots & 0       & 0      \\
 0      & 0      & t_3    & e_4    & \ldots & 0       & 0      \\
 \ldots & \ldots & \ldots & \ldots & \ldots & \ldots  & \ldots \\
 0      & 0      & 0      & 0      & \ldots & e_{N-1} & t_{N-1}\\
 0      & 0      & 0      & 0      & \ldots & t_{N-1} & e_N    \\
 \end{pmatrix}
\end{equation}
where $t_i$ is the hopping integral and $e_i$ is the electron
on-site energy. In the homogeneous lattice all diagonal
elements are equal and can be set to zero, $e_i = 0$.  The hopping integral is
written in the Su-Schrieffer-Heeger form \cite{Su_79, Su_80}:
\begin{equation}
  \label{c}
  t_i = - [v_0 - \alpha (x_{i+1} - x_i)] \,.
\end{equation}
This model mimics the synthetic DNA or peptide consisting of the
regular sequence of identical bases or peptide groups (see e.g.
\cite{Ari10}).

It is convenient to consider hamiltonian \eqref{a} in the
dimensionless quantities. Let us choose the  following units:
$v_0$ is the energy unit, $\sqrt{m \, k^{-1}}$  is the time unit
and $\sqrt{v_0 \,k^{-1}}$ is the length unit. Then the
dimensionless hamiltonian is
\begin{equation}
  \label{d}
  H = \dfrac12 \sum\limits_{i=1}^N \dot x_i +
      \dfrac12 \sum\limits_{i=1}^{N-1} q_i^2 -
      \sum\limits_{i=1}^{N-1}(1 - \alpha q_i) \,
      (\psi_i^* \psi_{i+1} + {\rm c.c.}) \,,
\end{equation}
where the same notation is preserved for the dimensionless
variables; $q_i \equiv (x_{i+1} - x_i)$ is the relative
displacement of the neighboring particles. The dimensionless
parameter $\alpha/\sqrt{k v_0} \to  \alpha$ is the free
parameter. Below we study the one-dimensional
polaron solutions of the hamiltonian \eqref{d}.

We find the solution for the standing stationary polaron when
all velocities in \eqref{d} $\dot x_i = 0$. The minimum of
energy \eqref{d} with respect to all lattice variables $\left\{
q_i \right\}$ and the electronic wave function $\left\{ \psi_i
\right\}$ determines the polaron.

If Eq. \eqref{d} is minimized with respect to $\{ \psi_i \}$
then $\vec\Psi$  is the eigenfunction of the electronic
hamiltonian $\widehat H^{\rm e}$ with energy $\varepsilon$. It
means that two procedures, -- energy minimization and solving
the stationary Schr\"odinger equation, are equivalent.

The electron wave function oscillates as $\psi_i e^{i
\varepsilon t/\hbar}$, where $\{ \psi_i \}$ are real amplitudes
of electron wave function. Note, the time oscillations are
eliminated in \eqref{d} because of complex conjugation. Then the
expression for the energy \eqref{d} is
\begin{equation}
  \label{e}
  H = \dfrac12 \sum\limits_{i=1}^{N-1} q_i^2 -
  2 \sum\limits_{i=1}^{N-1} (1-\alpha q_i) \psi_i \psi_{i+1} \,.
\end{equation}
The minimization \eqref{e} (i.e. $ \partial H / \partial q_i = 0 $) gives
\begin{equation}
  \label{f}
  q_i = -2 \alpha \psi_i \psi_{i+1}\,.
\end{equation}
Substitution of \eqref{f} into \eqref{e} results in:
\begin{equation}
  \label{g}
  H = - 2 \sum\limits_{i=1}^{N-1} \psi_i \psi_{i+1} -
  2 \alpha^2 \sum\limits_{i=1}^{N-1} (\psi_i \psi_{i+1})^2.
\end{equation}
An analogous approach was used in \cite{Con01}.

\begin{figure}
 \begin{center}
  \includegraphics[width=80 mm,angle=0]{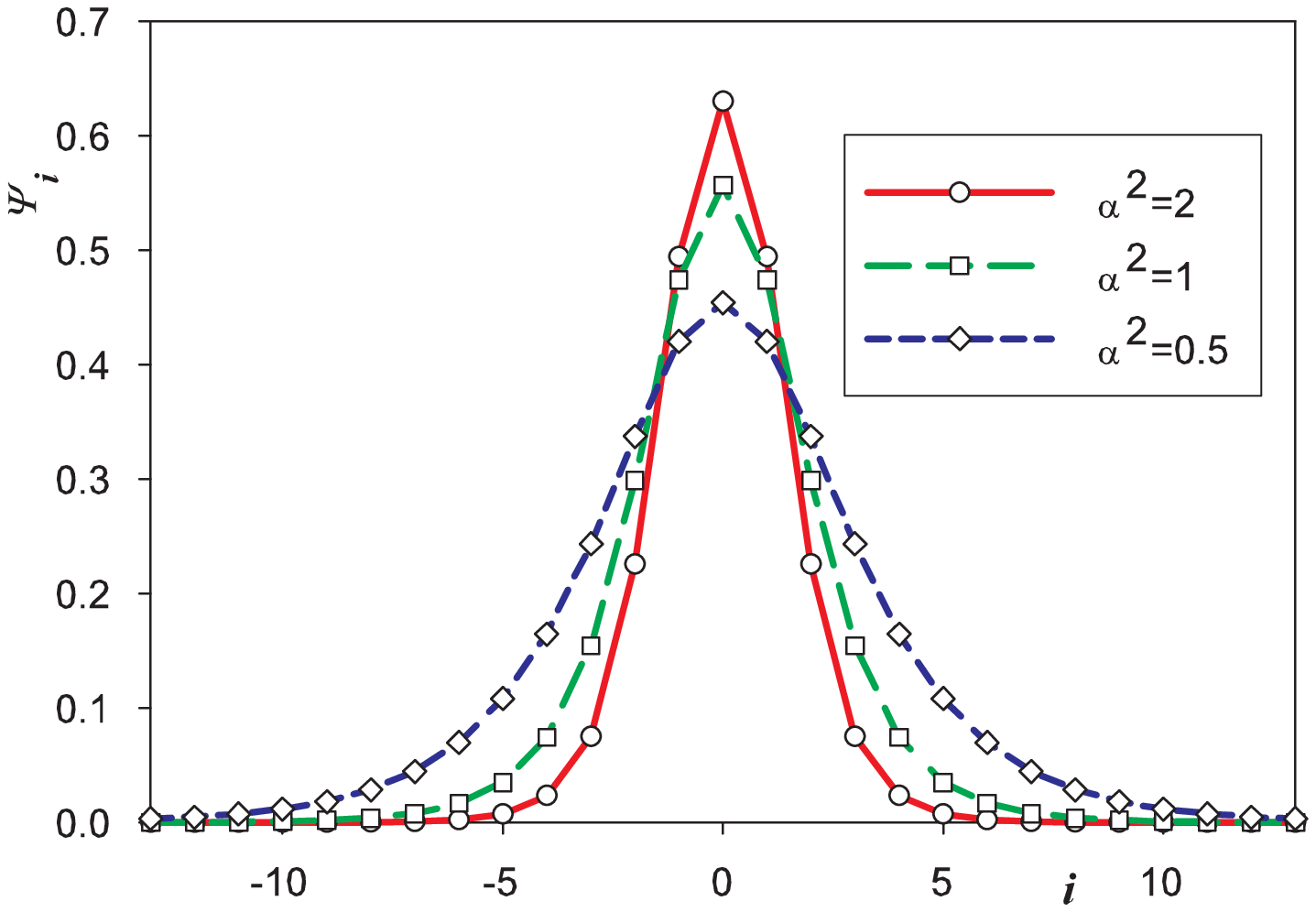}
  \includegraphics[width=80 mm,angle=0]{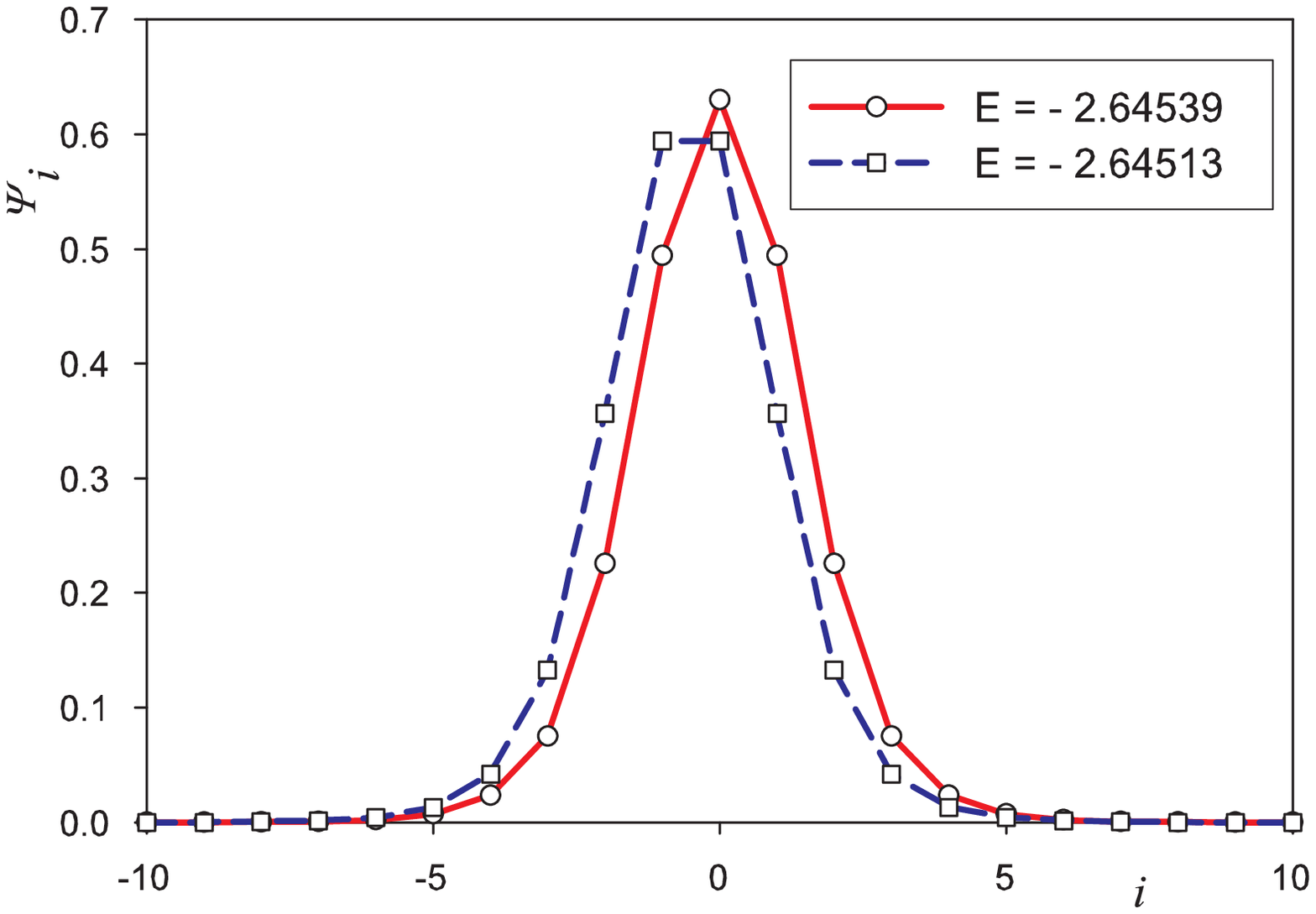}

 a \hspace{9 cm} b 
 \caption{
  \label{fig_1}
Results of the numerical calculations of the polaron.
 a) Wave functions for three values of parameter $\alpha$
(shown in insert). b) Site (solid line) and intersite (dashed
line) polarons for $\alpha^2 = 2$. Energies of these polarons are
shown in insert to the figure. Polarons are centered at the site
$i = 0$ ($-50 \le i \le 50$) in both figures.
       }
 \end{center}
\end{figure}

Expression \eqref{g} was minimized numerically preserving
the norm $\sum_i \psi_i^2 = 1$. If the lattice length $N$ is much
larger then the polaron radius then the polaron solution does not
depend on $N$. The numerical steepest descent method was used to
find the polaron profile on the lattice with $N = 101$. The
result is shown in Fig.~\ref{fig_1}a.

This polaron is centered on the particle. Such polarons will be
called the {\it site} polarons hereafter. The polaron center can
be located on the center of the bond between two nearest particles. Such polarons are
called {\it intersite} polarons. The energy is somewhat higher in
the latter case (see Fig.~\ref{fig_1}b).

%
%
%
%

Small energy difference of the site and intersite polarons can be
treated as the barrier separating two spatially different polaron
states with equal energies. The typical value for the energy unit
in DNA is $v_0 = 0{.}3$~eV. Then the energy difference between
site and intersite polarons is as small as $\lesssim 0.0001$~eV.
This small energy difference points to the possibility of
ballistically travelling polarons. This finding can
also be the starting point for contraction of
translationally-invariant states.

The total lattice contraction
$\Delta L = \sum_i q_i$, and using \eqref{f} one finds
\begin{equation}
  \label{h}
 \Delta L = - 2 \alpha \sum\limits_{i=1}^{N-1}\psi_i \psi_{i+1}.
\end{equation}
For large-width polaron $\psi_i \approx \psi_{i+1}$ and taking
into account the norm of the wave function (i.e. = 1) $\Delta L
\approx 2 \alpha$ .


\subsection{Analytical expressions for the small--radius polaron}

For definiteness we consider the site polarons hereafter. The
wave function of polaron is symmetrical relative to its center.
Consequently only one-half of a polaron (say, right half) can be
analyzed. This problem is considered on the semi-infinite
lattice and the new site numeration is used: $i = 0, 1,2,
\ldots, \infty$. The amplitudes of the wave function are denoted
as $\psi_0$ for the central polaron site and $\psi_i, \,\,
i=1,2, \ldots$ for other sites. Then the expression for energy
\eqref{g} is transformed to
\begin{equation}
  \label{i}
 H = -4 \sum\limits_{i=0}^{\infty} \psi_i \psi_{i+1}-
 4 \alpha^2\sum\limits_{i=0}^{\infty} (\psi_i \psi_{i+1})^2
\end{equation}
with the rescaled norm of the wave function
\begin{equation}
  \label{i1}
\psi_0^2 + 2 \sum_{i=1}^{\infty} \psi_i^2 = 1.
\end{equation}

\begin{figure}
 \begin{center}
  \includegraphics[width=80 mm,angle=0]{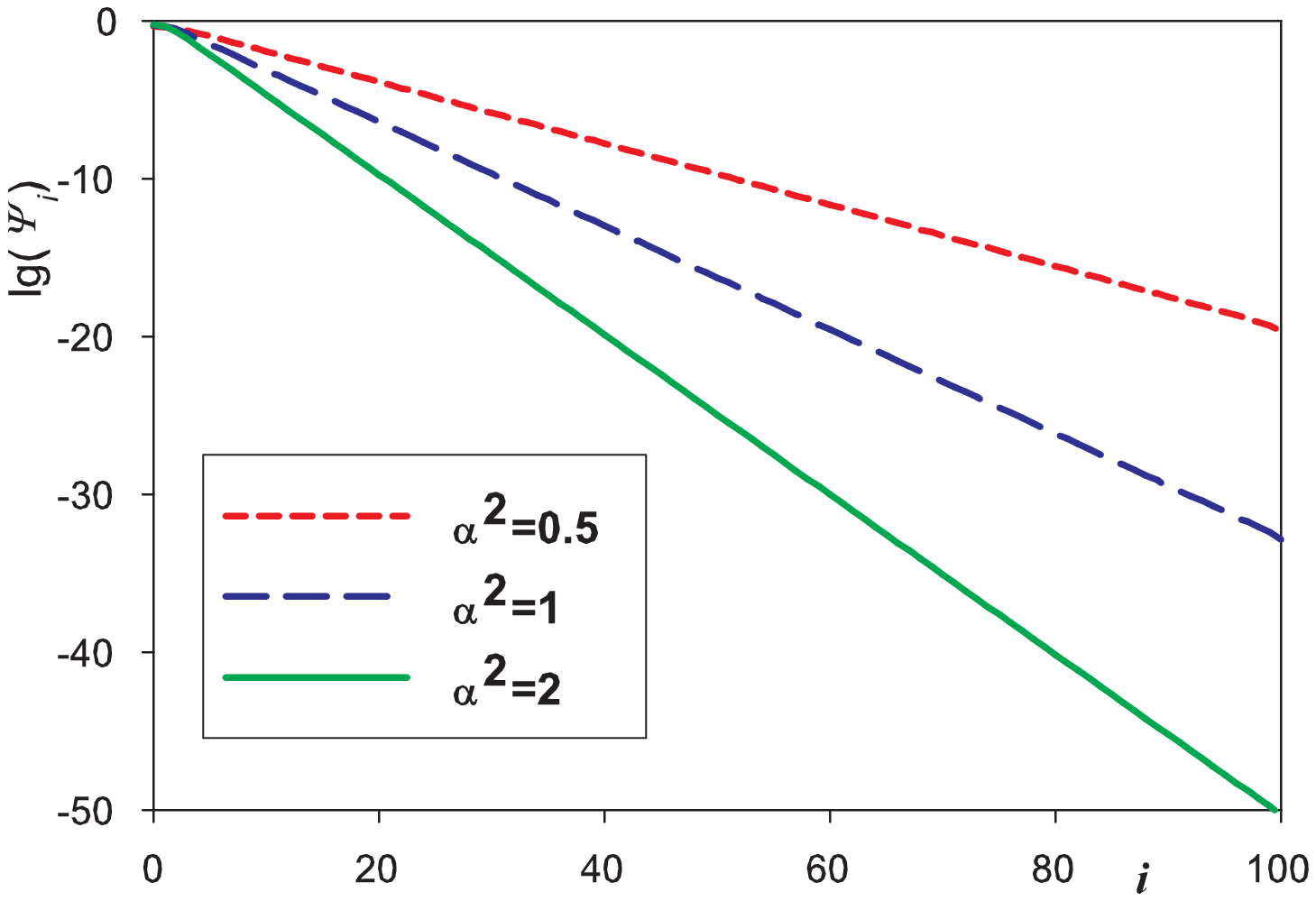}
  \includegraphics[width=80 mm,angle=0]{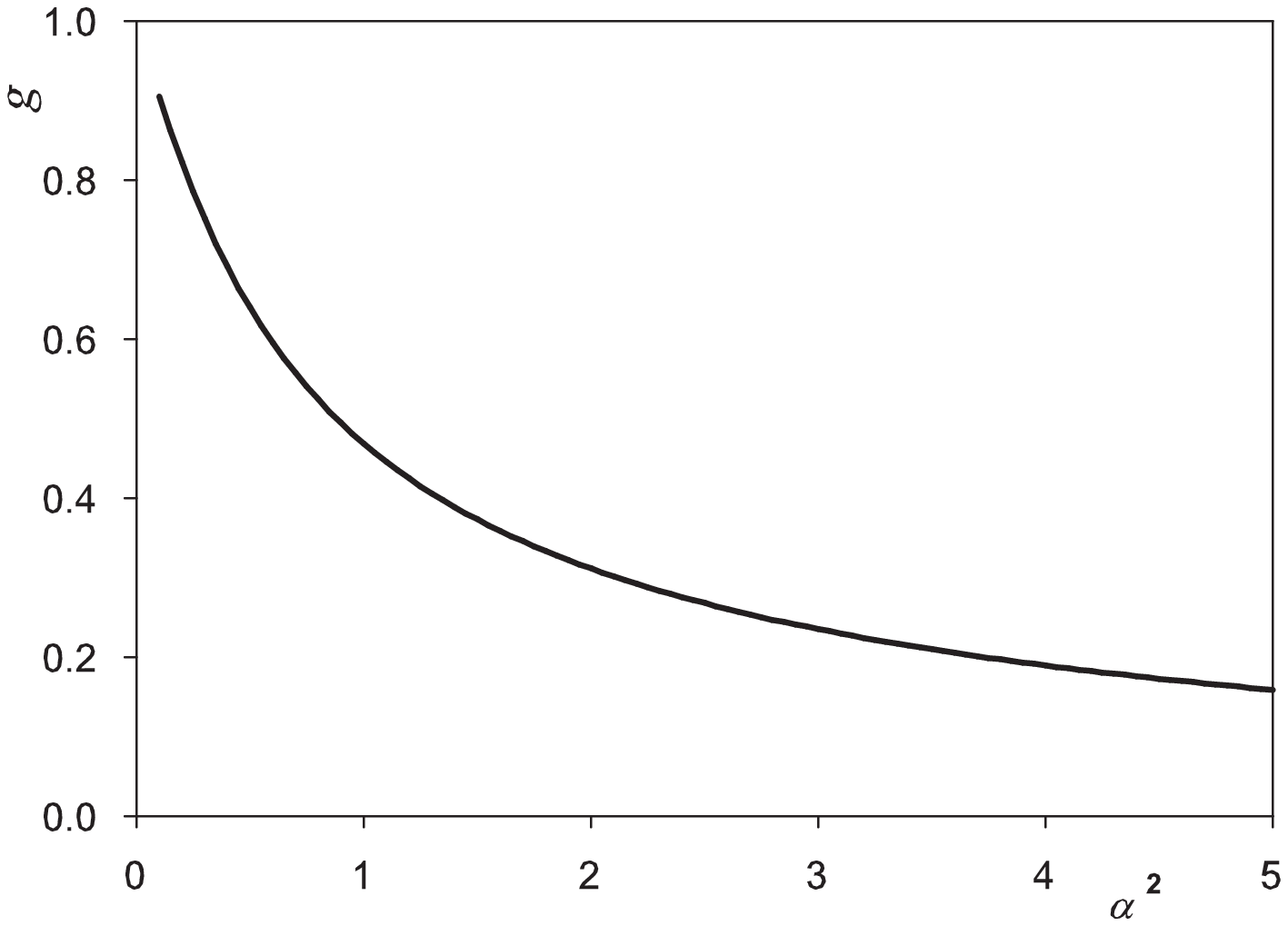}

 a \hspace{9 cm} b
  \caption{
  \label{fig_3}
a) Exponential decay of the wave function for different values
of $\alpha$. $N = 201$. b) The dependence of parameter $g$
{\it{vs.}} $\alpha$.
       }
 \end{center}
\end{figure}

Numerical minimization of energy \eqref{i} with respect to all
$\psi_i$ shows that the wave function decreases exponentially
with the growth of $i$ when $i \gg 1$ (see Fig.~\ref{fig_3}a),
i.e. $\psi_i \propto \exp(-i/R)$ and $R$ is the polaron radius.

It is convenient to introduce parameter $g$ related to the
radius $R$ by
\begin{equation}
  \label{el}
   R = \dfrac{1}{\ln g}\,.
\end{equation}
This relation follows from the exponential decay of the wave
function amplitude: $\psi_i \propto g^i$ at $i \gg
1$. In other words, $g$ is the common ratio of the geometrical
progression: $g = \psi_{i+1}/\psi_i$. Fig.~\ref{fig_3}b shows
the dependence of $g$ {\it{vs.}} parameter $\alpha$ of
electron-phonon interaction. One can see that the value of $g$
approaches unity for $\alpha \to 0$ and the polaron radius increases when parameter
$\alpha$ decreases.

The analytical solution can be constructed employing the
asymptotic behavior of the wave function at large $i$. Our goal
is to express the wave function and parameter $\alpha$ through
$g$.

It is convenient to introduce the Lagrange multiplier $\mu$ to
find the energy minimum of \eqref{i}. It allows to take into
account the condition for preserving the norm of wave function
\eqref{i1}. Then the minimization of the functional
\begin{equation}
  \label{j}
 \widetilde H = H + \mu \left( \psi_0^2 +
 2 \sum_{i=1}^{\infty} \psi_i^2 - 1  \right)
\end{equation}
with respect to $\psi_0, \psi_1, \psi_2, \ldots$ gives the
following system of equations
\begin{equation}
 \label{k}
\left\{
 \begin{split}
{} & \psi_1 + 2 \alpha^2 \psi_0 \psi_1^2=\dfrac{\mu}{2 \psi_0} \\
{} & \psi_{i+1} + \psi_{i-1} + 2 \alpha^2 \psi_i
\left(\psi_{i+1}^2 + \psi_{i-1}^2 \right) = \mu \psi_i;
 \qquad i \ge 1 \\
 \end{split}
\right.
\end{equation}
The Lagrange multiplier $\mu$ is expressed through $g$ as
\begin{equation}
  \label{m}
   \mu \approx g + \dfrac{1}{g}\,.
\end{equation}
This relation follows from the second equation \eqref{k} which
becomes linear for $i \gg 1$: $\psi_{i+1} + \psi_{i-1}
= \mu \psi_i$. The solution of this equation is  $\psi_i \propto
g^i$. It can be also verified that $\mu = - \varepsilon$ where
$\varepsilon$ is the electronic contribution to the polaron
energy.

The construction of the approximate analytical polaron solution
will be performed in the following manner. We suppose that the
exponential decay for the solution is a good approximation
starting from some site number $i_0$. It means that the cubic
terms in \eqref{k} can be neglected.

At the zeroth approximation the exponential decay starts from
$\psi_0$ ($i_0 = 0$) and $\psi_i = \psi_0 g^i$. This is very
rough approximation and we consider two next approximations.

\subsubsection{Polaron solution in the first approximation}

The exponential decay of the wave function starts from $i_0 = 1$
in the first approximation. And amplitudes are: $\psi_0, \psi_1,
\psi_2 = \psi_1 \, g, \psi_3 = \psi_1 \, g^2, \ldots$. Then only
two equations for $\psi_0$ and $\psi_1$ remain in \eqref{k}:
\begin{equation}
 \label{n}
\left\{
 \begin{split}
{} & \psi_1 + 2 \alpha^2 \psi_0 \psi_1^2=\dfrac{\mu}{2} \psi_0 \\
{} & \psi_1 g + \psi_0 + 2 \alpha^2 \psi_1
 \left( \psi_0^2 + g^2 \psi_1^2 \right) = \mu \psi_1 \,.
 \end{split}
\right.
\end{equation}
An expression for the norm of the wave function is
\begin{equation}
  \label{o}
   \psi_0^2 + 2 \dfrac{\psi_1^2}{1 - g^2} = 1
\end{equation}
and the second term in \eqref{o} is calculated as a sum of terms
of infinite geometrical progression.

As a result we get three algebraic equations \eqref{n} and
\eqref{o} for three variables $\psi_0 \,, \psi_1$ and $\alpha$. The exact
solution is too cumbersome and is not given. There exists a
limiting case when term $(g \psi_1)^2$ in \eqref{n} is
neglected. This approximation corresponds to the accuracy of the
first approximation. The answer in this case is very simple:
\begin{equation}
  \label{p}
   \psi_0 = \sqrt{\dfrac{1 - g^2}{2}}; \qquad
   \psi_1 = \dfrac{\sqrt{1 - g^4}}{2}.
\end{equation}
An expression for $\alpha$ is obtained by substitution \eqref{p}
and \eqref{m} into the first of equations \eqref{n}. The polaron
energy, expressed through $\psi_0, \psi_1$ and $\alpha$, is
\begin{equation}
  \label{q}
 E = -4 \left[
           \psi_0 \psi_1 + \dfrac{g \psi_1^2}{1- g^2} +
           \alpha^2 \psi_1^2
             \left(
                \psi_0^2 + \dfrac{g^2 \psi_1^2}{1 - g^4}
             \right)
        \right] \,.
\end{equation}

\begin{figure}
 \begin{center}
  \includegraphics[width=80 mm,angle=0]{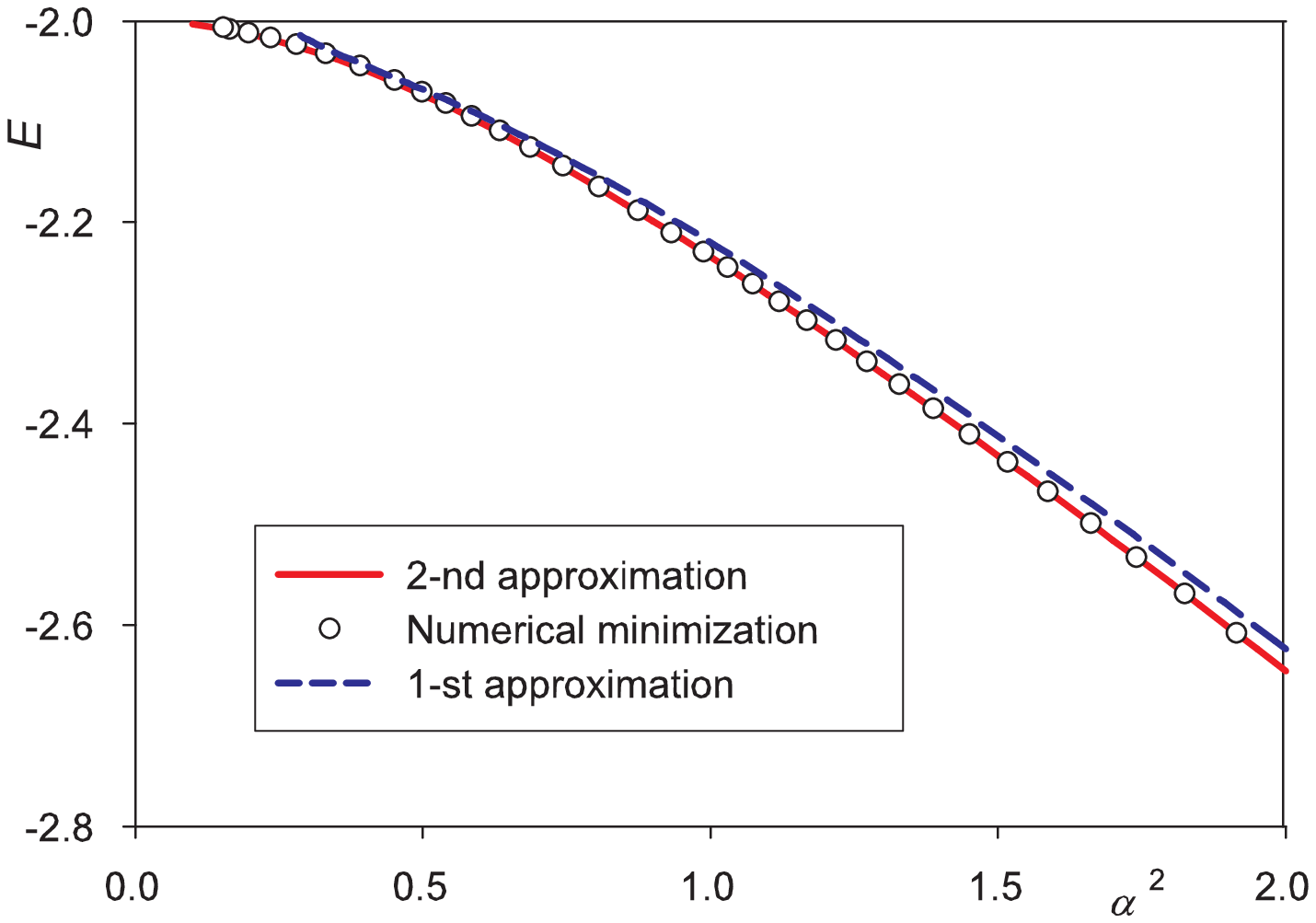}
  \includegraphics[width=80 mm,angle=0]{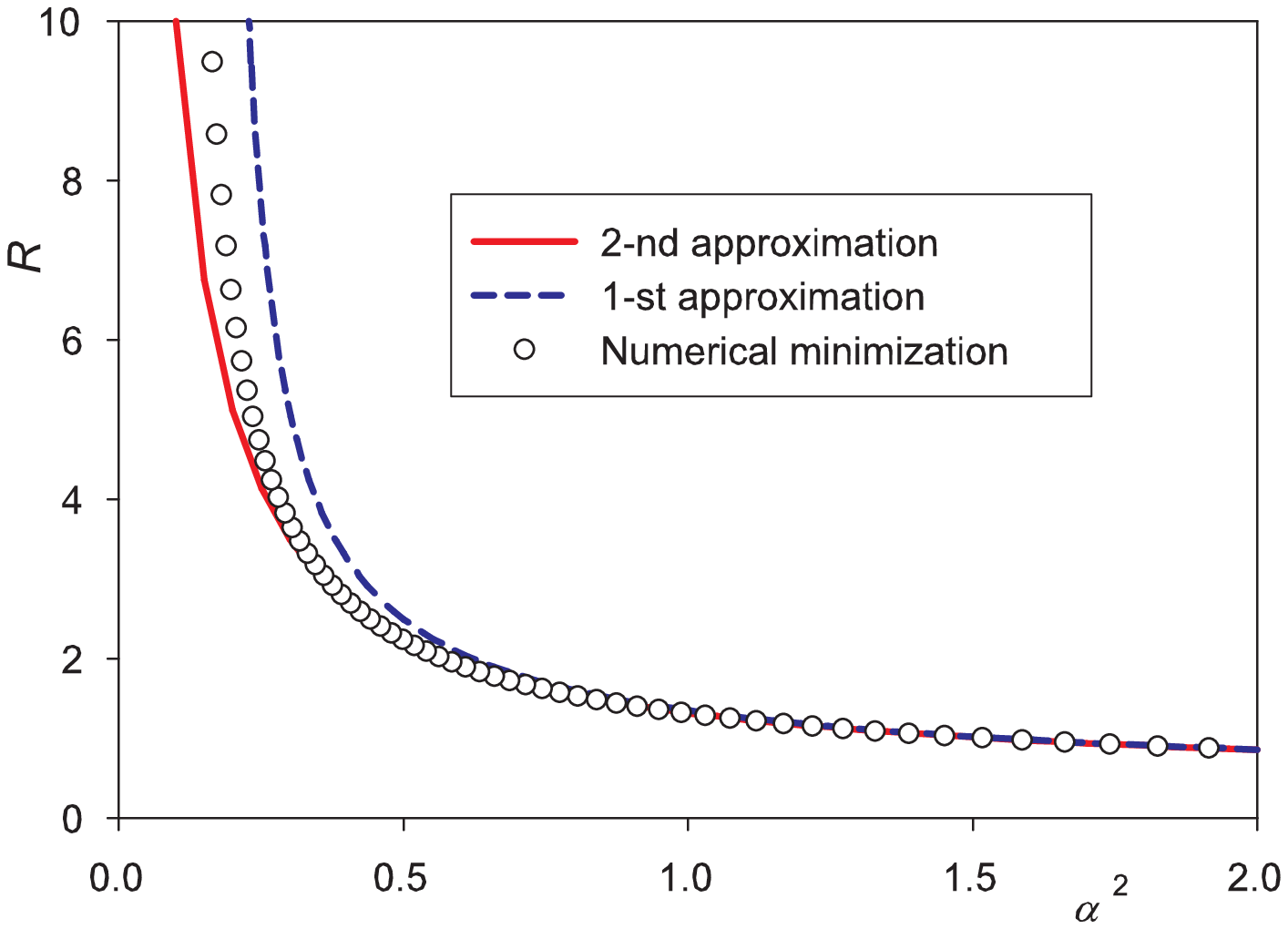}

\large a \hspace{9 cm} b

  \caption{
  \label{fig_5}
a) Polaron energy {\it{vs.}} parameter $\alpha$ of
electron-phonon interaction in the first and second
approximations. b) The polaron radius $R$ {\it{vs.}} parameter
$\alpha$. Results of numerical minimization are shown in empty
circles. $N=101$.
       }
 \end{center}
\end{figure}

The comparison of the results obtained in the first approximation
with the "exact" (numerical) results is shown in Fig.~\ref{fig_5}.
There is some discrepancy and the next approximation is
required.

%

\subsubsection{Polaron solution in the second approximation}

The exponential decay of amplitudes in the second approximation
starts from $i_0 = 2$. Amplitudes of wave function are: $\psi_0,
\psi_1, \psi_2, \psi_3 = \psi_2 \, g, \psi_4 = \psi_2 \, g^2,
\ldots$ and form the geometrical progression starting from the
third term. Three equations
\begin{equation}
 \label{r}
\left\{
 \begin{split}
{} & \psi_1 + 2 \alpha^2 \psi_0 \psi_1^2=\dfrac{\mu}{2} \psi_0 \\
{} & \psi_2 + \psi_0 + 2 \alpha^2 \psi_1
     \left(\psi_0^2 + \psi_2^2  \right) = \mu \psi_1 \\
{} & g \psi_2 + \psi_1 + 2 \alpha^2 \psi_2
     \left(\psi_1^2 + g^2 \psi_2^2 \right) = \mu \psi_2 \\
 \end{split}
\right.
\end{equation}
from system \eqref{k} is enough to solve the problem in the second approximation. The normalizing condition for the wave
function is
\begin{equation}
  \label{s}
 \psi_0^2 + 2 \psi_1^2 + 2 \dfrac{\psi_2^2}{1 - g^2} = 1 \,.
\end{equation}

Algebraic equations \eqref{r} and \eqref{s} are solved
numerically. The solution for the wave function is a series.
First three terms ($\psi_0, \psi_1, \psi_2$) are obtained from
the solution of these equations. Next terms ($\psi_i, i \ge 3$)
are members of the geometrical progression with the common ratio
$g$.

The comparison of "exact" (numerical minimization) and results
obtained in the second approximation is shown in
Figs.~\ref{fig_5}a and \ref{fig_5}b. One can see that the
coincidence is very good.

%
%
%


\section{The large--radius polaron}

The parameter $\alpha$ of electron-phonon interaction determines
the radius of polaron: polaron has large radius if $\alpha$ is
small. In this case the common ratio $g$ in the geometrical
progression is close to unity. We shall find the solution for
the large-radius polaron in the continuous approximation on the
infinite lattice $(- \infty < i < \infty)$.

The starting point for the evaluating the solution is the
same as in the case of small-radius polaron. Let us write this equation
once more for convenience:
\begin{equation}
  \label{3a}
  \psi_{i+1} + \psi_{i-1} + 2 \alpha^2 \psi_i
  \left( \psi_{i+1}^2 + \psi_{i-1}^2 \right) =
  \mu \psi_i \,.
\end{equation}
$g$ is slightly less then unity for the large-radius polaron,
and the Lagrange multiplier $\mu$  ($\mu = g + 1/g$) is somewhat
larger then 2. Lets introduce small parameter $\delta^2 \equiv
\mu -2$ and make some transformations in Eq.~\eqref{3a}.
Firstly, subtract $2 \psi_i$ from both sides of Eq.~\ref{3a}:
\begin{equation}
  \label{3b}
  \dfrac{1}{\delta^2}
  \left( \psi_{i+1} + \psi_{i-1} - 2 \psi_i \right) +
  2 \dfrac{\alpha^2}{\delta^2} \psi_i
\left( \psi_{i+1}^2 + \psi_{i-1}^2 \right) = \psi_i \,.
\end{equation}
Next, $\left( \psi_{i+1}^2 + \psi_{i-1}^2 \right)$  can be
substituted by $2 \psi_i^2$ because the difference between the
amplitudes of the wave function on neighboring sites is very
small for the large-radius polaron. As a result we get
\begin{equation}
  \label{3c}
  \dfrac{1}{\delta^2} (\psi_{i+1} + \psi_{i-1} - 2\psi_i) +
  4 \dfrac{\alpha^2}{\delta^2} \psi_i^3 = \psi_i \,.
\end{equation}
And, finally, we introduce new variables: $y_i =
\dfrac{\alpha \sqrt2}{\delta} \psi_i$. As a result the
following equation is obtained:
\begin{equation}
  \label{3d}
  \dfrac{1}{\delta^2} (y_{i+1} + y_{i-1} - 2y_i) +
 2 y_i^3 = y_i \,.
\end{equation}
First term in the left hand side of \eqref{3d} is the difference
form of the second derivative. Applying the continuous
approximation the following equation is obtained:
\begin{equation}
  \label{3e}
  y'' = y + 2 y^3 \,.
\end{equation}
Its solution is
\begin{equation}
  \label{3f}
  y = \dfrac{1}{\cosh(x)} \quad {\rm or} \quad
  \psi(x) = \dfrac{\delta}{\alpha \sqrt2}
  \dfrac{1}{\cosh(x)} \,.
\end{equation}
The normalization of the wave function gives $\delta =
\alpha^2$, then the final expression for the large-radius
polaron in discrete form is
\begin{equation}
  \label{3g}
  \psi_i = \dfrac{\alpha}{\sqrt2}
  \dfrac{1}{\cosh(\alpha^2 i)} \,.
\end{equation}
Very similar solution is obtained for a somewhat analogous
problem in \cite{Kal98}.

\begin{figure}
 \begin{center}
  \includegraphics[width=100 mm,angle=0]{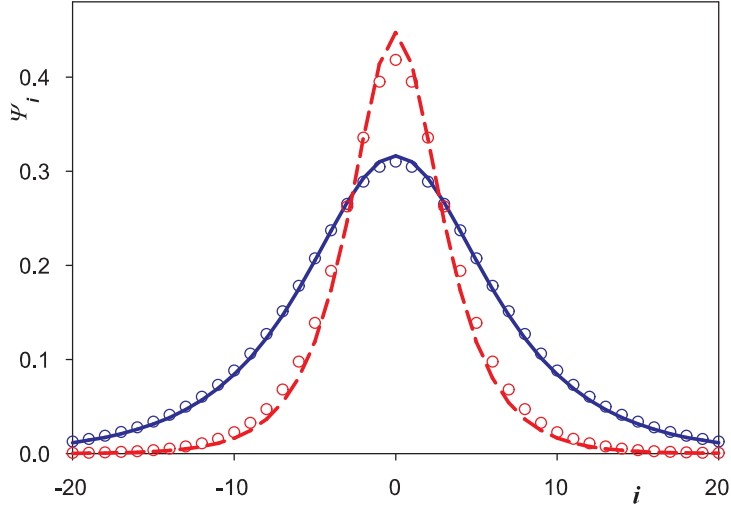}
  \caption{
  \label{fig_7}
The dependence of the amplitudes of wave function vs. the site
number $i$ according to \eqref{3g}. Dashed line -- for $\alpha =
0.4$, solid line -- for $\alpha = 0.2$. Circles -- accurate
results of numerical minimization.
}
 \end{center}
\end{figure}

The comparison of ``exact'' results (numerical minimization) and
the analytical expression \eqref{3g} is shown in
Fig.~\ref{fig_7} for two values of parameter $\alpha$. The
agreement is very good and it is the better the larger is
polaron radius. In one dimension case there is a transition from
the small to the large polaron and $\alpha \approx 0{.}6$
separates these two solutions.


\section{The dynamics of polaron formation}

Two cases of large and small polarons were considered above.
These solutions correspond to the energy minimum in the
stationary state. And now we consider the evolution of the
dynamical and electronic degrees of freedom giving the polaron
as the final state.

The following numerical experiment was performed. The wave
function with some profile was chosen as the initial electronic
condition. Initial conditions for the dynamical variables:
lattice at rest, i.e. $x_i = 0, \, \dot x_i = 0$.

The system evolution is determined by the following system
\begin{equation}
 \label{4a}
\left\{
 \begin{split}
\ddot x_i = & (x_{i-1} + x_{i+1} - 2 x_i)
            +  \alpha [
            (\psi_i \psi_{i+1}^* + \psi_i^* \psi_{i+1})  -
            (\psi_{i-1} \psi_i^* + \psi_{i-1}^* \psi_i)]  \\
\dot \psi_i = & \dfrac{i}{\widetilde h}
    \left[
    (1 - \alpha q_i) \psi_{i-1} + (1 - \alpha q_i) \psi_{i+1}
    \right]
 \end{split}
\right.
\end{equation}
where $q_i \equiv (x_{i+1} - x_i)$ and $\widetilde h$ is the
dimensionless Planck's constant: $\dfrac{\hbar}{v_0}
\sqrt{\dfrac{m}{k}} \to \widetilde h$. If DNA parameters are
chosen \cite{Con00} then $1/\widetilde h \approx 80$.

If the wave function is initially located on one site $\left(
\psi_i(t=0) \propto \delta_{i,i_0} \right)$, or has the constant
value on the lattice $\left( \psi_i(t=0) \propto N^{-1/2}
\right)$, then the wave function spreads over lattice very
rapidly and polarons are not formed. The reason is that the
characteristic electronic time is very small $(\sim \widetilde h
\approx 10^{-2})$ compared to the dynamical time scale $(\sim
1)$. And the dynamical degrees of freedom have no enough time to
follow the wave function.

\begin{figure}
 \begin{center}
  \includegraphics[width=100 mm,angle=0]{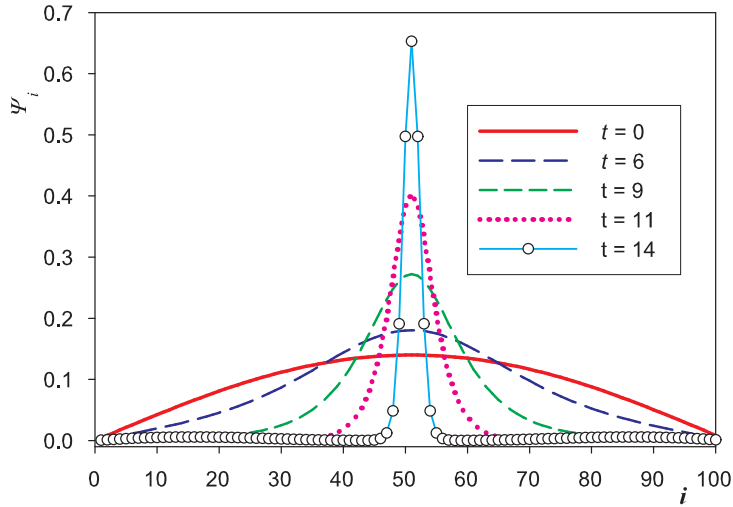}
  \caption{
  \label{fig_9}
Snapshots of the wave function evolution at different time
instances shown in insert. Initial wave function $\psi_i(t=0)
\propto \left( \sin[\pi i/(N+1)] , 0 \right)$. $\alpha^2 = 2, N =
101$. At $t = 14$ the polaron is completely formed.
         }
 \end{center}
\end{figure}

If the initial wave function is chosen as a eigenfunction of
unperturbed lattice, then the lattice ``captures'' the
oscillating electron and the joint electron--lattice evolution
results in polaron formation. The trial complex wave function is
chosen in the form $\psi_i^0 (t = 0) = \left(
\sqrt{\dfrac{2}{N+1}} \sin \left[ \dfrac{\pi i}{N+1} \right],0
\right)$. It corresponds to the ground-state wave function of
the unperturbed lattice. The evolution of the wave function is
shown in Fig.~\ref{fig_9}. The polaron is formed during few
oscillation periods of the lattice.

Eqs~\eqref{4a} correspond to the microcanonical ensemble where
total energy is preserved. The energy gain due to the polaron
formation is compensated by the vibrational excitation of a
lattice. Because of the dynamical interaction with the lattice,
polaron fluctuates, slightly changes its form and position, but
stays very stable. If frictional forces at the lattices ends
$-\gamma \dot x_1$ and $-\gamma \dot x_N$ are introduced then all
vibrations decay and only polaron survives on the lattice.

If the initial wave function is the eigenfunction of the first
excited state $\psi_i^1 (t = 0) = \left( \sqrt{\dfrac{2}{N+1}}
\sin \left[ \dfrac{2 \pi i}{N+1} \right], 0 \right) $ then {\it
two} polarons are formed but with norm of wave function = 0.5
for each of them.

The problem of charge transfer from a donor to the DNA double
strand is very complex quantum-dynamical problem. An accurate
solution of of time dependent Schr\"odinger equation for coupled
state of excited donor and a lattice is necessary. We did not
touch this problem here and limited ourself by the model
representation of the initial wave functions.


\section{Wave function evolution on the finite-sized lattice}

Polarons on the infinite lattice were considered in the previous
sections. But if the particular applications of the suggested
approach are of interest (e.g. charge transfer in DNA) then the
lattice should have the finite length, i.e. $N \lesssim 100$.
And there appearers few peculiarities which differ the polaron
dynamics on the infinite lattice from the lattice of final
length. Few examples are considered below.

First of all, the intriguing property of the evolution equations
\eqref{4a} should be mentioned. At some initial conditions the
interaction of an electron with the lattice is the ``one-way'':
There exists an influence of the lattice on the electron through
electron-phonon interaction, but the back ``polaronic'' action
of an electron on the lattice is absent. In other words,
additional forces acting on particles from the electron
subsystem are zeroes. In the simplest case it is realized  if
the wave function is initially localized on one (arbitrary)
site. Let the initial wave function for definiteness is real and
totally located on the first site: $\psi_1(t=0) = 1$. From the
second set of equations \eqref{4a} it follows that at the
certain time step the wave function on odd sites is real, and on
even sites is imaginary. At the next time step the situation
inverts:
 the wave function on odd sites is imaginary, and on even sites is real.
 Then the products $\psi_i \psi_{i+1}^*$ is always zero (see \eqref{4a}). Hence, the
lattice always stays to be harmonic. This conclusion is also valid
if the initial wave function is organized in such a manner that
it is real on all even sites and imaginary on all odd sites (or
vise verse).

\begin{figure}
 \begin{center}
  \includegraphics[width=80 mm,angle=0]{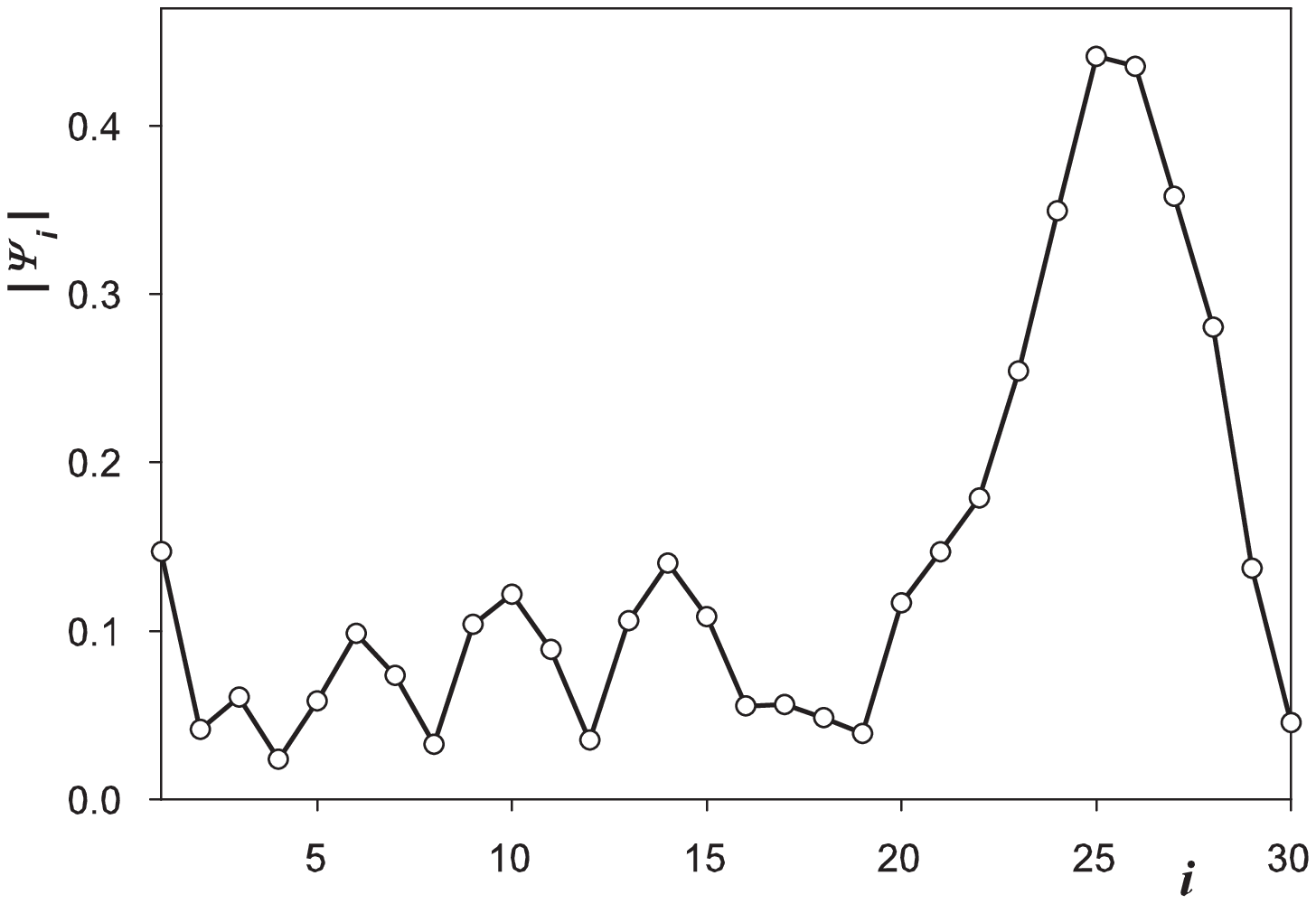}
  \includegraphics[width=80 mm,angle=0]{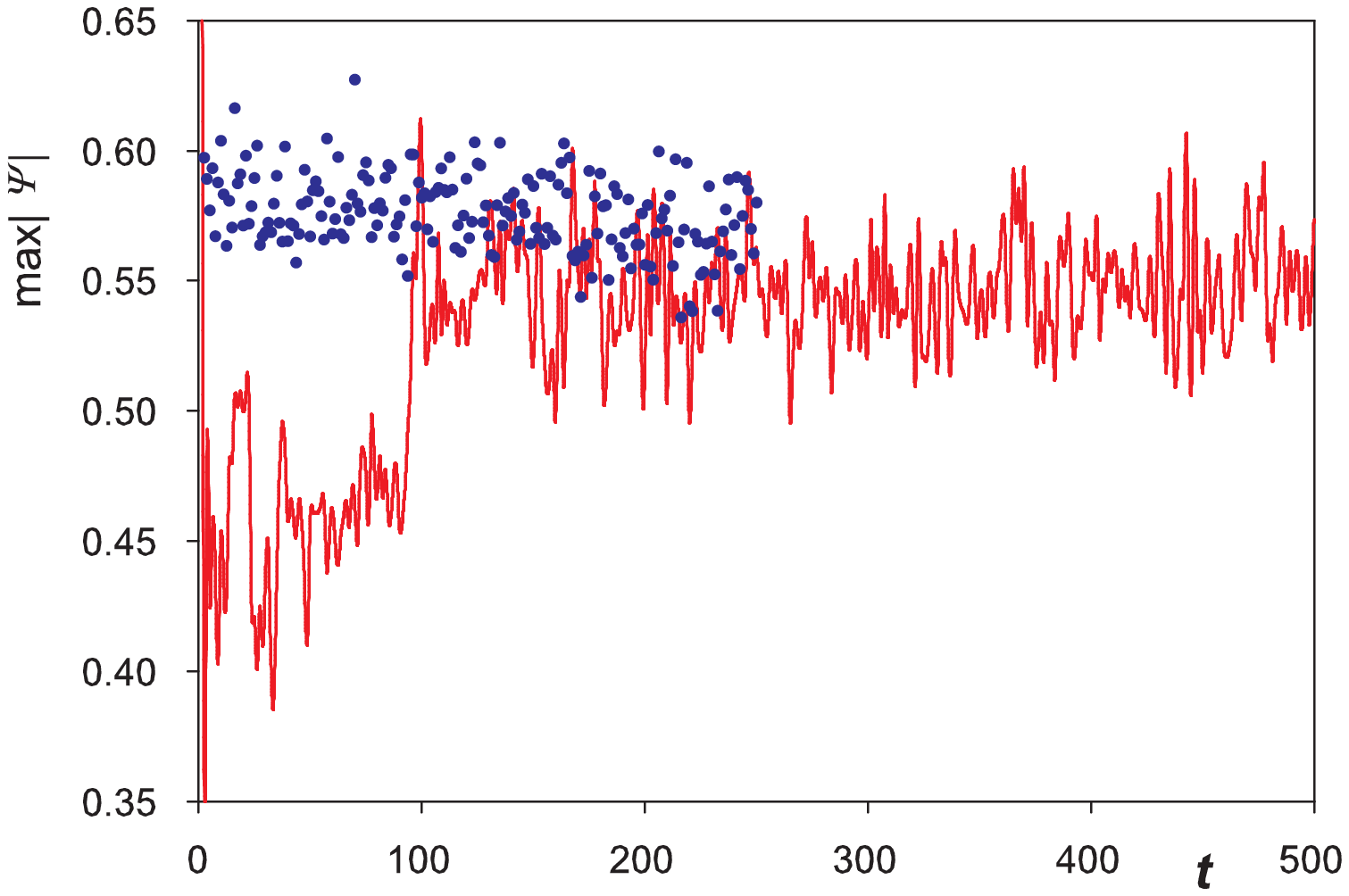}
  \caption{
  \label{fig_10}
a) Snapshot of the polaron wave function at  time $t = 134$.
$\alpha^2 = 2, N = 30, \widetilde e_{25} = 1.5, \widetilde
\alpha_{25} = 0.1$. b) Maximal value of the wave function (in
absolute value) {\it{vs.}} time.  Solid line for
$\widetilde\alpha = 0.1$, dots for $\widetilde\alpha = 1$. Every
point is the result of averagin on the time interval $t = 1.25$.
$N = 30, \alpha^2 = 2$.
         }
 \end{center}
\end{figure}

Second interesting case is the wave function evolution on the
inhomogeneous lattice. Let one (arbitrary) lattice site $i_0$
has the TBA parameters which differs from others. These
parameters are $\widetilde e_{i_0} \neq 0$ for the electron
on-site interaction, and $\widetilde \alpha \neq \alpha$ for the
electron-phonon interaction \eqref{c}. Initial wave function can
be chosen arbitrary. If the wave function for the homogeneous
lattice ($\widetilde e = 0, \widetilde \alpha = \alpha$) never
forms the localized state -- polaron, but for the lattice with
defect the case is opposite: the polaron is formed after some
induction period. The result is shown in Fig.~\ref{fig_10}a.
Though the norm of the polaron wave function is less then unity,
polaron stays a stable quasiparticle for a very long time.

The polaron formation can be observed as follows: The maximal
value of the wave function amplitude is monitored and this
maximum is averaged over some time interval. The formation of
the stable maximum with max$|\psi| = 0.5-0.6$ is the direct
evidence of the polaron formation (see Fig.~\ref{fig_10}b).
Polaron is formed practically instantaneously at
$\widetilde\alpha = 1$, and after induction period $t \approx
100$ for $\widetilde\alpha = 0.1$.
%
%
%

\begin{figure}
 \begin{center}
  \includegraphics[width=90 mm,angle=0]{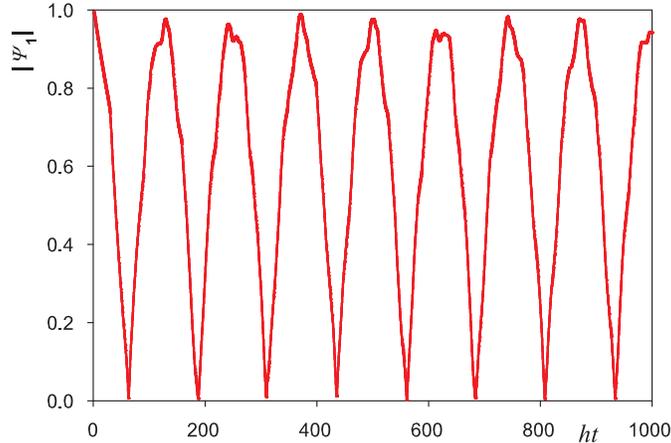}
  \caption{
  \label{fig_12}
Absolute value of the amplitude of the wave function on the
first site. $N = 30, \widetilde t_1 = 0.1$.
         }
 \end{center}
\end{figure}

There exists one more interesting case of the wave function
evolution at some special choice of parameters. If parameter
$\widetilde e = 0$ (absence of the diagonal disorder) and only
one hopping integral differs from the others, then it is the
case when the lattice does not influence the electron evolution
(as discussed above). The usual ``hopping'' dynamics on the
homogeneous lattice will be observed. Let only first site
differs from the others by the hopping constant $\widetilde t_1
\neq t$ (see \eqref{b}) and the wave function is localized on
the site $i = 1$ at the initial time $t=0$. Naturally, no
polaron is formed as the lattice does not influence the electron
dynamics. But there exists the phenomenon of the periodic
returning of the wave function to the first site. This
phenomenon is related with the final lattice size and is shown
in Fig.~\ref{fig_12}.
Time is measured in ``electronic'' time units $\widetilde h t$
which are related to dynamical time units by the dimensionless
Planck's constant $\widetilde h = 0.0125$ (see above).

This problem (in very simple formulation) reminds experiments on
the photoinduced charge transfer from intercalated
metallocomplex to the DNA chain \cite{Bar11}, and the value
$\widetilde t_1$ reflects the interaction of the ligand with the
DNA bases.

Some interesting effects are experimentally observed:
``chemistry at a distance'' \cite{Gen10} and ``ping-pong
electron transfer \cite{Eli08}. Our results can help to
elucidate some experimental results which has no satisfactory
interpretation yet \cite{Gen10a}. Particularly the detailed
analysis of the ``electronic ping-pong'' will be given in our
next paper where moving polarons will be considered.

The polaron behavior and wave function evolution were considered
at zero temperature till now. The kinetic energy of thermal
fluctuation can be rather high at ambient temperatures, and the
polaronic time of life can be limited due to interactions with
these fluctuations. Below we consider the evolution of the
wave function at finite temperatures.

\begin{figure}
 \begin{center}
  \includegraphics[width=90 mm,angle=0]{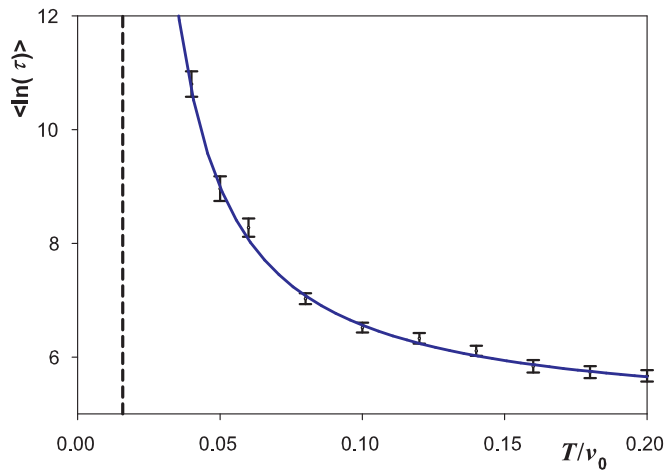}
  \caption{
  \label{fig_13}
Polaron time of life $\tau$ vs. the temperature. Temperature is
measured in units of energy ($v_0 = 0.3$~eV for DNA). Solid line
is the hyperbolic fitting $\lg(\tau) = a + b/(T - T_0)$, where
$a,b$ are some parameters, and $T_0$ is the ``critical''
temperature. Dashed line shows the asymptotic value $T_0 \approx
0.016$. Error bars correspond averaging over 20 random
trajectories.
         }
 \end{center}
\end{figure}

The following problem was solved numerically to find out the
role of temperature. The stationary polaron was formed at the
lattice center, as described in the previous section. Then the
lattice was thermalized from its ends (Langevine random forces
with friction) to achieve the necessary temperature. If the
temperature is high enough, the polaron is destroyed rather
soon. We monitor the maximal value of the squared absolute value
of the wave function amplitude on the lattice, computing its
averaged value over time interval $t=50$. In the beginning this
value is close to $\sim0.4$, what approximately corresponds to
the squared maximum of amplitude for the standing polaron. Then
this maximum falls down to $\sim0.1$ at some time $\tau$. This
time corresponds to the polaron life time. When temperature
decreases, $\tau$ increases. Life time $\tau$ {\it vs.}
temperature is shown in Fig.~\ref{fig_13}

Of course, there no any firm basis to make a conclusion that the
found temperature $T_0$ is the true critical temperature. Sooner
this temperature should be determined by the lattice length $N$
and the parameter $\alpha$ of electron-phonon interaction. It
seems reasonable that the "critical" temperature should
decrease with the growth of the lattice length $N$ and decreasing
parameter $\alpha$.

\section{Conclusions}

Analytical expressions for the large- and small-radius polarons
are derived in the present paper. The polaron radius is
determined by the parameter $\alpha$ of the electron-phonon
interaction in the SSH model: polaron is small at $\alpha
\gtrsim 0{.}6$. The solution for the wave function of
small-radius polaron represents a series where the first three
terms are the solution of the system of algebraic equations and
the next terms are geometrical progression. This form of
solution is explained by the exponential decay of the wave
function with increasing distance from the polaron center:
$\psi_{i+1}/\psi_i = g = $ const. For the large-radius polaron
an analytical expression is derived in the continuous
approximation. This solution reminds the solutions of some
nonlinear equations (nonlinear Schr\"odinger, modified Korteweg
-- de Vries). Good agreement between analytical solutions and
numerical simulation is obtained.

The considered simple model can help to explain existing
experiments on the long range coherent charge transfer in DNA.


\end{document}